\documentclass[11pt]{article}
\usepackage{moriond,epsfig}

\def\Journal#1#2#3#4{{#1} {\bf #2}, #3 (#4)}

\def\NCA{{\em Nuovo Cimento} A}
\def\NCC{{\em Nuovo Cimento} C}

\def\NIMA{{\em Nucl. Instrum. Methods} A}
\def\NPA{{\em Nucl. Phys.} A}
\def\NPB{{\em Nucl. Phys.} B}
\def\PLB{{\em Phys. Lett.}  A}
\def\PLB{{\em Phys. Lett.}  B}
\def\PRL{\em Phys. Rev. Lett.}
\def\PRC{{\em Phys. Rev.} C}
\def\PRD{{\em Phys. Rev.} D}

\def\APP{\em Astrop. Phys.}
\def\NJP{\em New Journal of Physics}
\def\EPJC{{\em Eur. Phys. J.} C}
\def\EPJdC{{\em Eur. Phys. J. direct} C}

\def\be{\begin{equation}}
\def\ee{\end{equation}}
\def\bea{\begin{eqnarray}}
\def\eea{\end{eqnarray}}

\newcommand{\lsim}{\mathrel{\mathop{\kern 0pt \rlap
  {\raise.2ex\hbox{$<$}}}
  \lower.9ex\hbox{\kern-.190em $\sim$}}}
\begin{document}
\vspace*{1cm}

\begin{flushright}
{\bf Contributed paper to the Rencontres de Moriond \hspace{0.1cm} $\,$ \\}
{\bf "Electroweak Interactions and Unified Theories", \hspace{0.1cm} $\,$ \\}
{\bf La Thuile, Aosta Valley, Italy, March 2004 \hspace{0.1cm} $\,$ \\} 
\end{flushright}
\vspace{1.5cm} 

\title{DAMA/NaI RESULTS}
\author{R. BERNABEI$^1$, P. BELLI$^1$, F. CAPPELLA$^1$,
F. MONTECCHIA$^{1,}$\footnote{also: \uppercase{U}niversit\`a "\uppercase{C}ampus
\uppercase{B}io-\uppercase{M}edico"
di \uppercase{R}oma, 00155, \uppercase{R}ome, \uppercase{I}taly},
F. NOZZOLI$^1$, A. INCICCHITTI$^2$, D. PROSPERI$^2$, R. CERULLI$^3$,
C.J. DAI$^4$, H.H. KUANG$^4$, J.M. MA$^4$, Z.P. YE$^{4,}$\footnote{also: 
\uppercase{U}niversity of \uppercase{Z}hao
\uppercase{Q}ing, \uppercase{G}uang \uppercase{D}ong, \uppercase{C}hina}}

\address{$^1$Dipartimento di Fisica, Universit\`a di Roma ``Tor Vergata'' \\
and INFN, Sezione di Roma2, I-00133 Rome, Italy \\
$^2$Dipartimento di Fisica, Universit\`a di Roma ``La Sapienza'' \\
and INFN, Sezione di Roma, I-00185 Rome, Italy \\
$^3$INFN - Laboratori Nazionali del Gran Sasso, I-67010 Assergi (Aq), Italy \\
$^4$IHEP, Chinese Academy, P.O. Box 918/3, Beijing 100039, China}

\maketitle\abstracts{
The DAMA/NaI set-up of the DAMA experiment has been operative during seven annual cycles
and has investigated several rare processes. In particular, it has been realised in order  
to investigate the model independent annual modulation signature for Dark Matter 
particles in the galactic halo. With the total exposure collected in the 
seven annual cycles (107731 kg $\times$ day) 
a model independent evidence for the presence of a Dark Matter particle component
in the galactic halo has been pointed out at 6.3 $\sigma$ C.L.. Some of the many
possible corollary model dependent quests for the candidate
particle have been presented as well.}

\section{Introduction}

DAMA is an observatory for rare processes based on the 
development and use of
various kinds of radiopure scintillators.
It has realised several low background set-ups;
the main ones are:
i) DAMA/NaI ($\simeq$ 100 kg of
radiopure NaI(Tl)), which was put out of operation in July 2002 
\cite{Nim98,RNC,Psd96,Diu,Nairp,Mod1,Mod2,Ext,Mod3,Sist,Sisd,Inel,Hep}; 
ii) DAMA/LXe ($\simeq$ 6.5 kg liquid Xenon) \cite{2};
iii) DAMA/R\&D, devoted to tests on prototypes and small scale experiments 
\cite{3}; 
iv) the new second generation set-up 
DAMA/LIBRA ($\simeq$ 250 kg;
more radiopure NaI(Tl)) in operation since March 2003.

The results obtained by DAMA/NaI, investigating over seven annual cycles
(107731 kg $\times$ day total exposure) the presence of a Dark Matter particle 
component in the
galactic halo by means of the model independent WIMP annual modulation signature,
have been 
published in ref.\cite{RNC} together 
with some of the many possible corollary model dependent quests for the candidate
particle. 
We invite the reader to refer to ref.\cite{RNC} for a discussion of 
experimental and theoretical arguments related to the results and to model dependent comparisons
and - more in general - to our literature to distinguish what DAMA has presented 
(and its meaning) with respect with what sometimes is quoted by others.  

We remind that the annual modulation signature \cite{Freese} is very distinctive
since it requires the simultaneous satisfaction of
all the following requirements: the rate must contain a component
modulated according to a cosine function (1) with 
one year period, $T$, (2)
and a phase, $t_0$, that peaks around $\simeq$ 2$^{nd}$ June (3);
this modulation must only be found
in a well-defined low energy range, where WIMP induced recoils
can be present (4); it must apply to those events in
which just one detector of many actually "fires" 
({\it single-hit} events), since
the WIMP multi-scattering probability is negligible (5);
 the modulation
amplitude in the region of maximal sensitivity is expected 
to be $\lsim$7$\%$  (6). This latter rough limit 
would be larger in case of other possible scenarios such as 
e.g. those in refs. \cite{Wei01,Fre03}.
To mimic such a signature 
spurious effects or side reactions
should be
able both to account for the whole observed modulation amplitude 
and to contemporaneously satisfy all the requirements; 
no one has been found or suggested 
by anyone over about a decade.

The DAMA/NaI set-up and its performances have been described 
in ref.\cite{Nim98} and some other information on its performances and
upgrading can be found in refs.\cite{RNC,Sist}. 
Here, we just remind that
the two PMTs, coupled through 10 cm long Tetrasil-B light guides to each NaI(Tl) crystal, 
worked in coincidence with hardware thresholds at single photoelectron 
level in order to assure
high efficiency for the coincidence at few keV level \cite{Nim98}.
The energy threshold of the experiment, 2 keV, 
was instead determined by means of X ray sources
and of keV range Compton electrons on the basis also of 
the features of the noise rejection procedure
and of the efficiencies when lowering the number 
of available photoelectrons \cite{Nim98}. 

\section{A brief description of the final model independent result over 7 annual cycles}

\begin{figure}[!ht]
\begin{center}
\includegraphics[height=10.5cm]{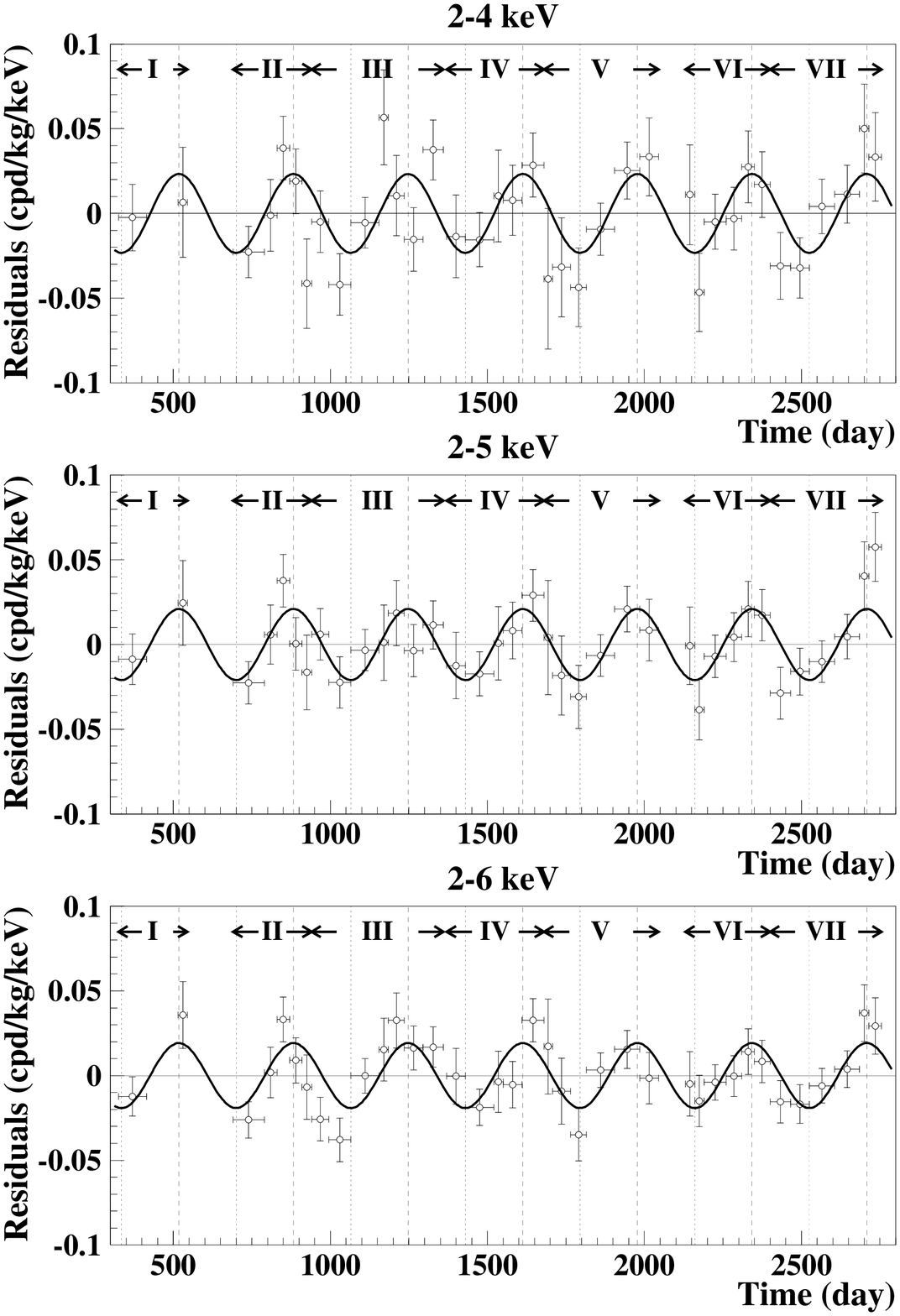}
\includegraphics[height=6.5cm]{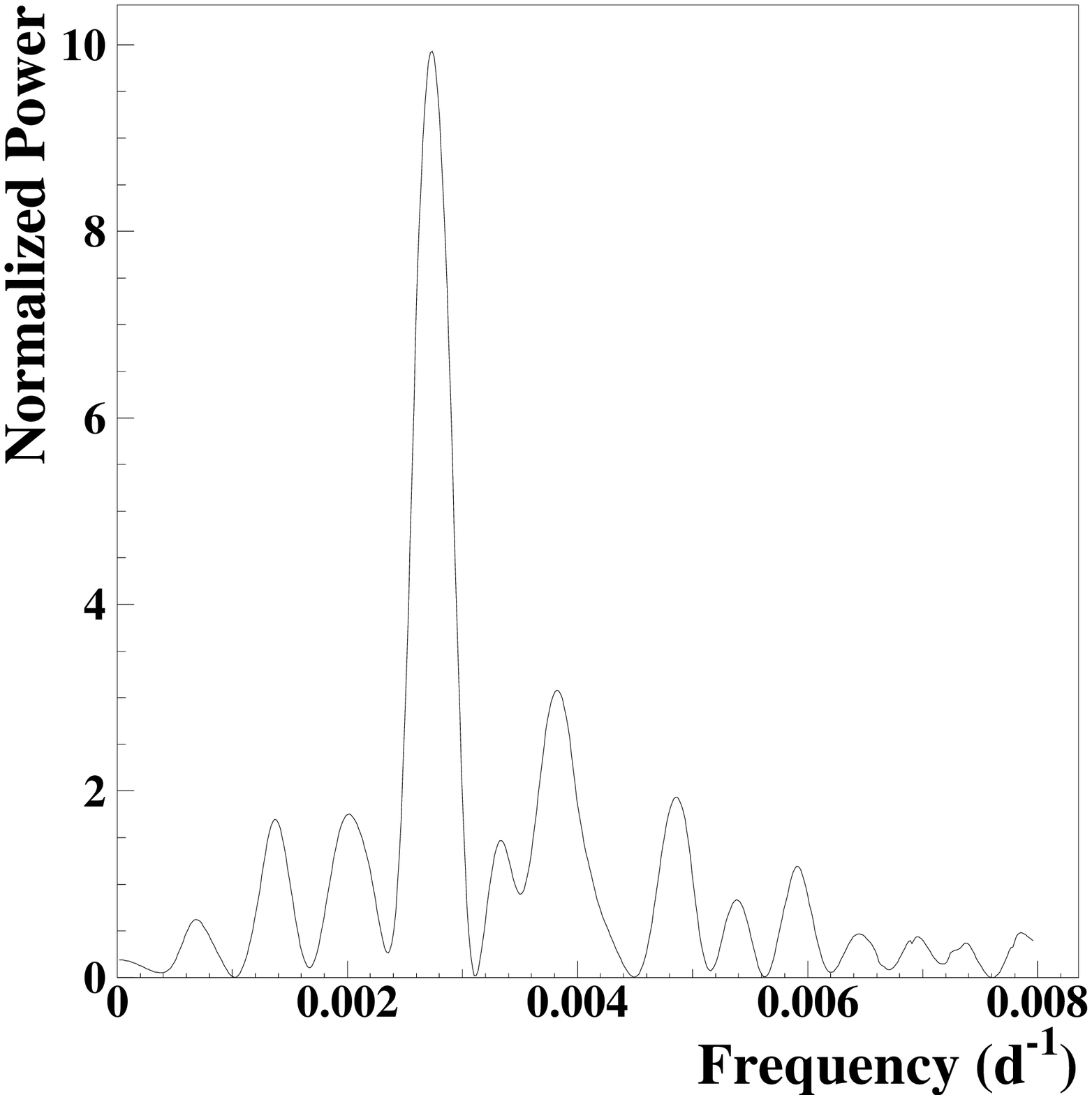}
\end{center}
\caption{{\it On the left}: experimental 
residual rate for {\it single-hit} events in the (2--4), (2--5) and
(2--6) keV energy intervals
as a function of the time over 7 annual cycles (total exposure 
107731 kg $\times$ day); end of data taking July 2002.
The experimental points present the errors as vertical
bars and the associated time bin width as horizontal bars. The 
superimposed curves represent the cosinusoidal function
behaviours expected for a WIMP signal 
with a period equal to 1 year and phase exactly at $2^{nd}$ June; the modulation
amplitudes have been obtained by best fit. See also ref. $^2$. {\it On the right}:
power spectrum of the measured (2--6) keV {\it single-hit} residuals 
calculated including also the treatment of the experimental
errors and of the time binning.
As it can be seen, the principal mode corresponds to a
frequency of $2.737 \cdot 10^{-3}$ d$^{-1}$, 
that is to a period of $\simeq$ 1 year.}
\label{fig1}
\end{figure}

A model independent approach on the data of the 
seven annual cycles offers an immediate evidence of the presence of 
an annual modulation of the rate of the {\it single-hit} events
in the lowest energy region as shown in the residuals of 
Fig. \ref{fig1} -- {\em left},
where the time behaviours of the measured (2-4), (2-5) and (2-6) keV
{\it single-hit} events residual rates are reported.
The data favour the presence of a modulated cosine-like behaviour 
($A \cdot$ cos$\omega (t-t_0)$) at 6.3 $\sigma$ C.L. 
and their fit for the (2--6) keV 
larger statistics energy interval offers modulation amplitude equal to
$(0.0200 \pm
0.0032)$ cpd/kg/keV,
$t_0 = (140 \pm 22)$ days and
$T = \frac{2\pi}{\omega} = (1.00 \pm 0.01)$ year, 
all parameters kept free in the fit.
The period and phase agree with those expected in the case of a
WIMP induced effect ($T$ = 1 year and $t_0$ roughly at 
$\simeq$ 152.5$^{th}$ day of the year). 
The $\chi^2$ test on the (2--6) keV residual rate 
disfavours the hypothesis of unmodulated behaviour giving a 
probability of $7 \cdot
10^{-4}$ ($\chi^2/d.o.f.$ = 71/37).
The same data have also been investigated by a Fourier analysis
as shown in Fig.~\ref{fig1} -- {\em right}.
Modulation is not observed above 6 keV \cite{RNC}.
Finally, a suitable statistical analysis has shown that the 
modulation amplitudes are statistically well distributed 
in all the crystals,
in all the data taking periods and  
considered energy bins. More arguments can be found in ref. \cite{RNC}.
A careful investigation of all the known possible sources
of systematic and side
reactions has been regularly carried out and published at time of each data 
release while detailed quantitative discussions can be found 
in refs. \cite{RNC,Sist} \footnote{
We take this opportunity just to remind that the
experimental set-up, located deep underground, 
was equipped with a neutron shield made by Cd foils and 
polyethylene/paraffin moderator; moreover, a $\simeq$ 1 m of concrete almost completely
surrounds the installation acting as a further neutron moderator. Note that
the sizeable discussion reported in
\cite{RNC,Sist} already demonstrates that
a possible modulation of the neutron flux
(e.g. possibly observed by the ICARUS coll., as reported in the ICARUS internal report TM03-01,
and expected in a MonteCarlo
simulation by H. Wulandari et al. as reported in hep-ex/0312050)
cannot quantitatively
contribute to the DAMA/NaI observed modulation amplitude, even if the neutron flux would
be assumed to be 100 times larger than that measured at LNGS by various authors
over more than 15 years. In addition, 
it cannot satisfy all the peculiarities of the signature mentioned above.
Finally, we also remind that the contribution of solar neutrinos, whose flux is also
expected to be modulated, is many orders of magnitude lower than
the measured rate (see e.g. Astrop. Phys. 4 (1995) 45).}.
No systematic effect or side reaction able to account for the 
observed modulation amplitude and to mimic 
a WIMP induced effect has been found. 
\begin{figure}[!ht]
\centering
\includegraphics[height=6.cm]{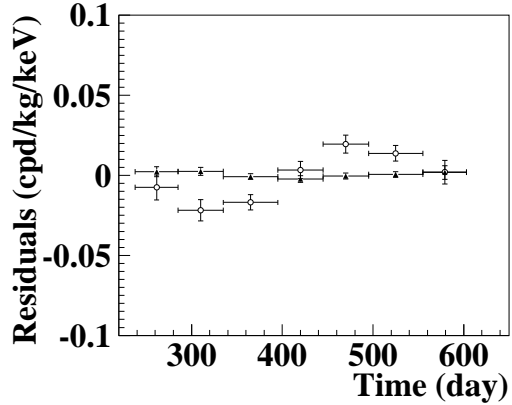}
\caption{Experimental residual rates over seven annual cycles
for {\it single-hit} events (open circles) -- class of events 
to which WIMP events belong --
and over the last two annual cycles for {\it multiple-hits} 
events (filled triangles)
-- class of events to which WIMP events do not belong -- in the
(2--6) keV cumulative energy interval. They have been obtained 
by considering for each class of events the data as collected 
in a single annual cycle and using in both cases the same 
identical hardware and the same identical software
procedures. The initial time is taken on August 7$^{th}$.
See text.}
\label{fig2}
\end{figure}
As a further relevant investigation, 
the {\it multiple-hits} events also collected during the DAMA/NaI-6 
and 7 running periods
(when each detector was equipped with
its own Transient Digitizer with a dedicated renewed electronics)
have been studied and 
analysed by using the same identical hardware and the same 
identical software 
procedures as for the case of the {\it single-hit} events (see Fig. \ref{fig2}).
The {\it multiple-hits} events class -- on the contrary of the {\it single-hit} one
-- does not include events induced by
WIMPs since the probability that a WIMP scatters 
off more than one detector is
negligible.  
The fitted modulation amplitudes are: $A=(0.0195\pm0.0031)$ cpd/kg/keV 
and $A=-(3.9\pm7.9) \cdot 10^{-4}$ cpd/kg/keV for {\it single-hit} 
and {\it multiple-hits} residual rates, respectively.
Thus, evidence of annual modulation is present in 
the {\it single-hit} residuals (events class to which the WIMP-induced recoils
belong), while it is absent in the 
{\it multiple-hits} residual rate (event class to which only background events belong).
Since the same identical hardware and the same identical software procedures have been 
used to analyse the two classes of events,
the obtained result offers an additional strong support for the presence of 
Dark Matter particles in the galactic halo further excluding any side effect 
either from hardware or from software procedures or from background.
 
\vspace{0.4cm}
In conclusion, the presence of an annual modulation in the
residual rate of the {\it single-hit} events 
in the lowest energy interval (2 -- 6) keV,
satisfying all the features expected for a WIMP component  
in the galactic halo
is supported by the data of the seven annual cycles at 6.3 $\sigma$ C.L..
This is the experimental result of DAMA/NaI. It is model
independent; no other experiment whose result can be
directly compared with this one is available so far in the field of Dark Matter
investigation.

\section{A brief description of some of the possible corollary model dependent quests for a candidate}

On the basis of the obtained model independent result, corollary 
investigations can also be
pursued on the nature and coupling of the WIMP candidate.
\begin{figure}[!ht]
\begin{center}
\includegraphics[height=10.cm]{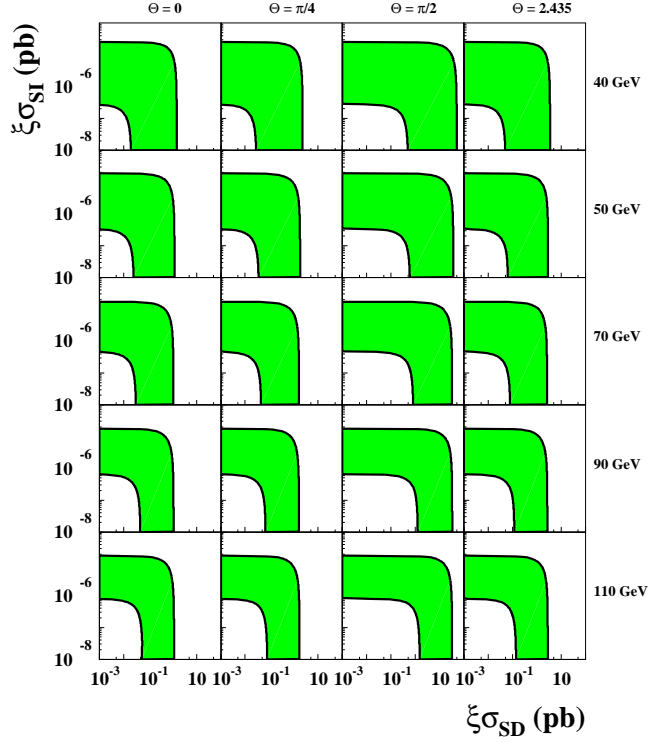}
\end{center}  
\caption{{\it Case of a WIMP with mixed SI\&SD interaction for the model frameworks given in ref. $^2$.}
Coloured areas: example of slices (of
the allowed volume) in the plane 
$\xi \sigma_{SI}$ vs
$\xi \sigma_{SD}$ for some of the possible $m_W$ and $\theta$ values.
Inclusion of other existing uncertainties on parameters and models 
would further extend the regions; for example,
the use of more favourable form factors and/or of more favourable
spin factors than the ones considered here 
would move them towards lower cross sections.
For details see ref.$^2$.}
\label{fig3}
\end{figure}
This latter investigation is instead model
dependent and -- considering the large uncertainties which exist on the 
astrophysical, nuclear and particle physics
assumptions and on the parameters needed in the calculations -- has no general
meaning (as it is also the case of exclusion plots, of expected recoil 
energy behaviours and 
of the WIMP parameters evaluated in indirect
search experiments). Thus, it should be
handled in the most general way as we have preliminarily
pointed out with time passing in the past
\cite{Mod1,Mod2,Ext,Mod3,Sist,Sisd,Inel,Hep}.
Some specific details can be found in ref. \cite{RNC}
and references therein, with devoted discussions
of experimental and theoretical aspects and evaluations. 
Here we only remind that the results summarised here are not exhaustive 
of the many scenarios possible at present level of knowledge, including 
those depicted in some more recent works such as e.g. refs. \cite{Fre03,Kam03}.

\begin{figure}[!ht]
\begin{center}
\includegraphics[height=6.cm]{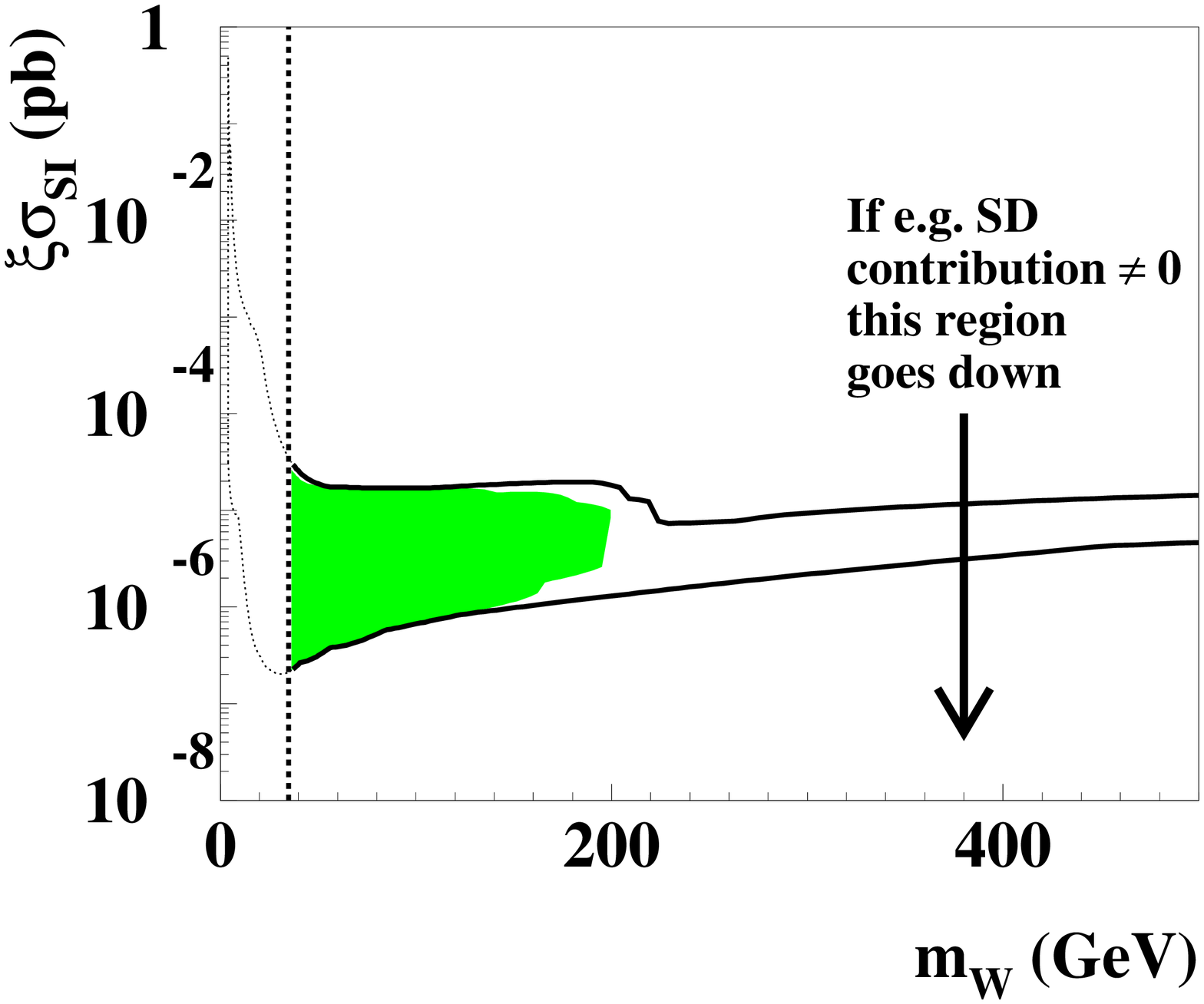}
\includegraphics[height=6.cm]{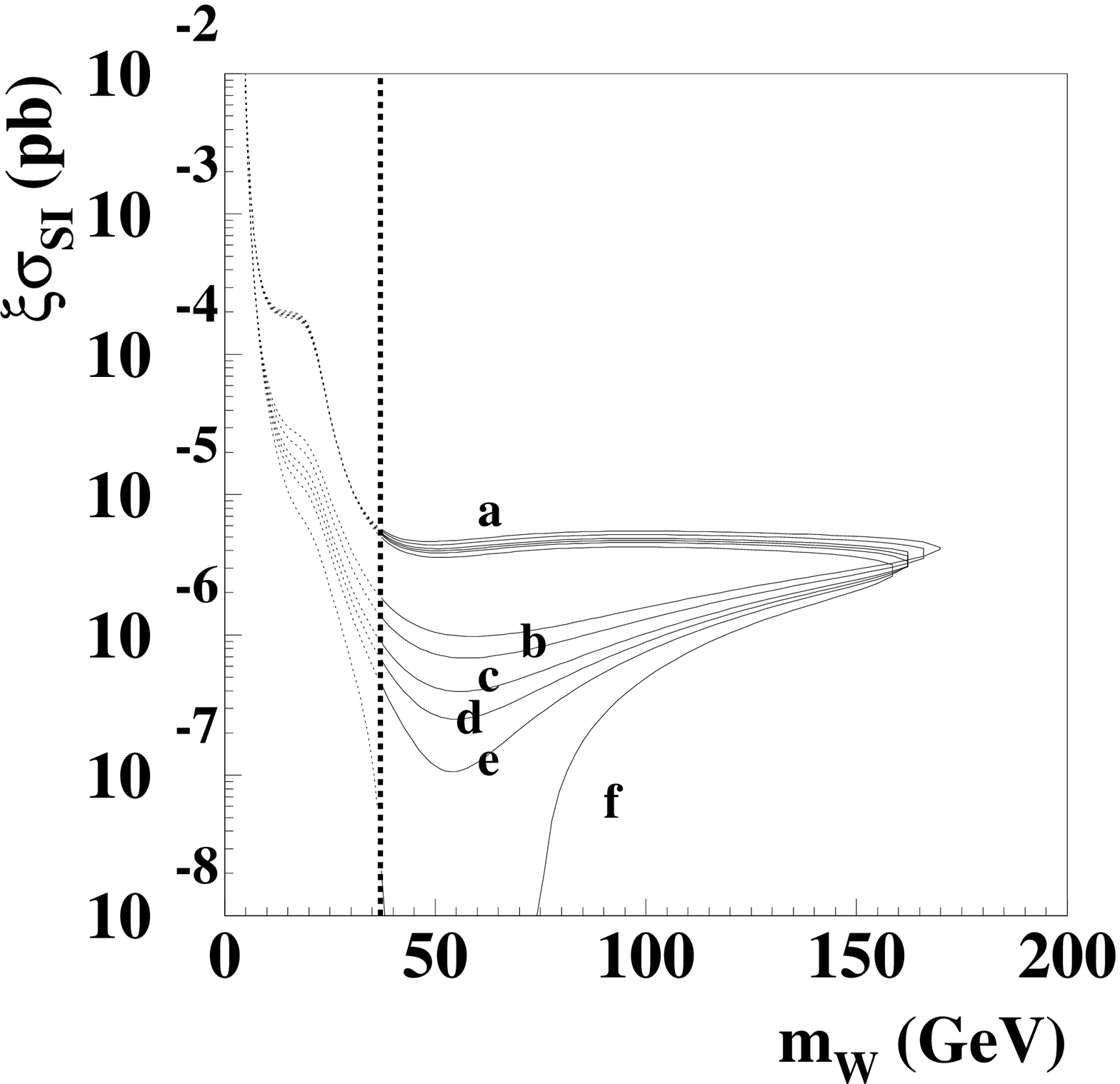}
\end{center}  
\caption{{\it On the left : Case of a WIMP with dominant SI 
interaction for the model frameworks given in ref.$^2$.} 
Region allowed in the plane ($m_W$, $\xi
\sigma_{SI}$). The vertical dotted line represents a bound in case
of a neutralino candidate when
supersymmetric schemes based on GUT assumptions
are adopted to analyse the LEP data;
the low mass region is allowed for neutralino when other schemes are
considered (see e.g. refs. $^{20,21,22}$) and for every other WIMP candidate.
While the area at WIMP masses above 200 GeV is allowed 
only for few configurations, the lower one is allowed by most configurations
(the colored region gathers only those above
the vertical line). 
The inclusion of other existing uncertainties on parameters and models
would further extend the region; for example,
the use of more favourable SI form factor for Iodine
alone would move it towards lower cross sections.
{\it On the right:
Example of the effect induced by the inclusion of a SD component
different from zero on
allowed regions given in the plane $\xi\sigma_{SI}$ vs $m_W$.}
In this example the
Evans' logarithmic axisymmetric $C2$ halo model with
$v_0 = 170$ km/s, $\rho_0$ equal to the maximum value for this model
and a given set of the parameters' values (see ref. $^2$) have been considered.
The different regions refer to different SD contributions for the particular case of
$\theta = 0$:
$\sigma_{SD}=$ 0 pb (a), 0.02 pb (b), 0.04 pb (c), 0.05 pb (d),
0.06 pb (e), 0.08 pb (f). Analogous situation is found for the other
model frameworks. For details see ref. $^2$.}
\label{fig4}
\end{figure}

\begin{figure}[!ht]
\begin{center}
\includegraphics[height=6.cm]{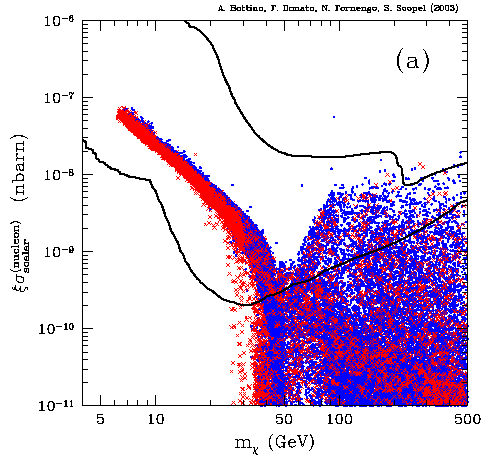}
\end{center}  
\caption{Theoretical expectations of $\xi \sigma_{SI}$ versus $m_W$
in the purely SI coupling for the particular case of a neutralino candidate
in the MSSM with gaugino mass unification at GUT scale; the curve is the same 
as in Fig. 4-{\it left}. Figure taken from ref. $^{20}$.}
\label{fig5}
\end{figure}

\begin{figure}[!ht]
\begin{center}
\includegraphics[height=6.cm]{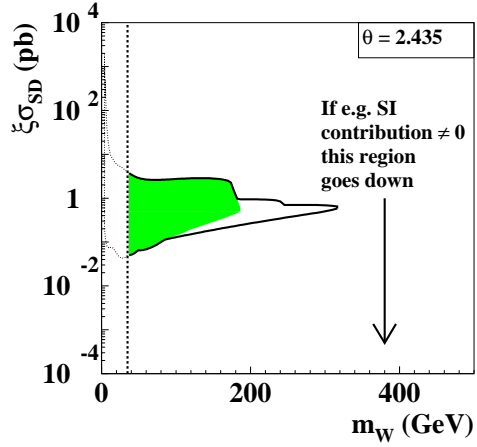}
\end{center}  
\caption{{\it Case of a WIMP with dominant SD interaction in the model
frameworks given in ref. $^2$.} An example of
regions allowed in the plane ($m_W$, $\xi \sigma_{SD}$) at given
$\theta$ value ($\theta$ is defined in the $[0,\pi)$ range);
here $\theta = 2.435$ ($Z_0$ coupling). 
For the definition of the vertical line and of the coloured area see 
the caption of Fig. 4. 
Inclusion of other existing uncertainties on
parameters and models (as discussed in ref. $^2$)
would further extend the SD allowed regions. For example,
the use of more favourable SD form factors
and/or more favourable spin factors
would move them towards lower cross sections.
Values of $\xi \sigma_{SD}$ lower than those corresponding to this allowed
region are possible also e.g. in case of an even small
SI contribution (see ref. $^2$). For details see ref.$^2$.}
\label{fig6}
\end{figure}

\begin{figure}[!ht]
\begin{center}
\includegraphics[height=5.5cm]{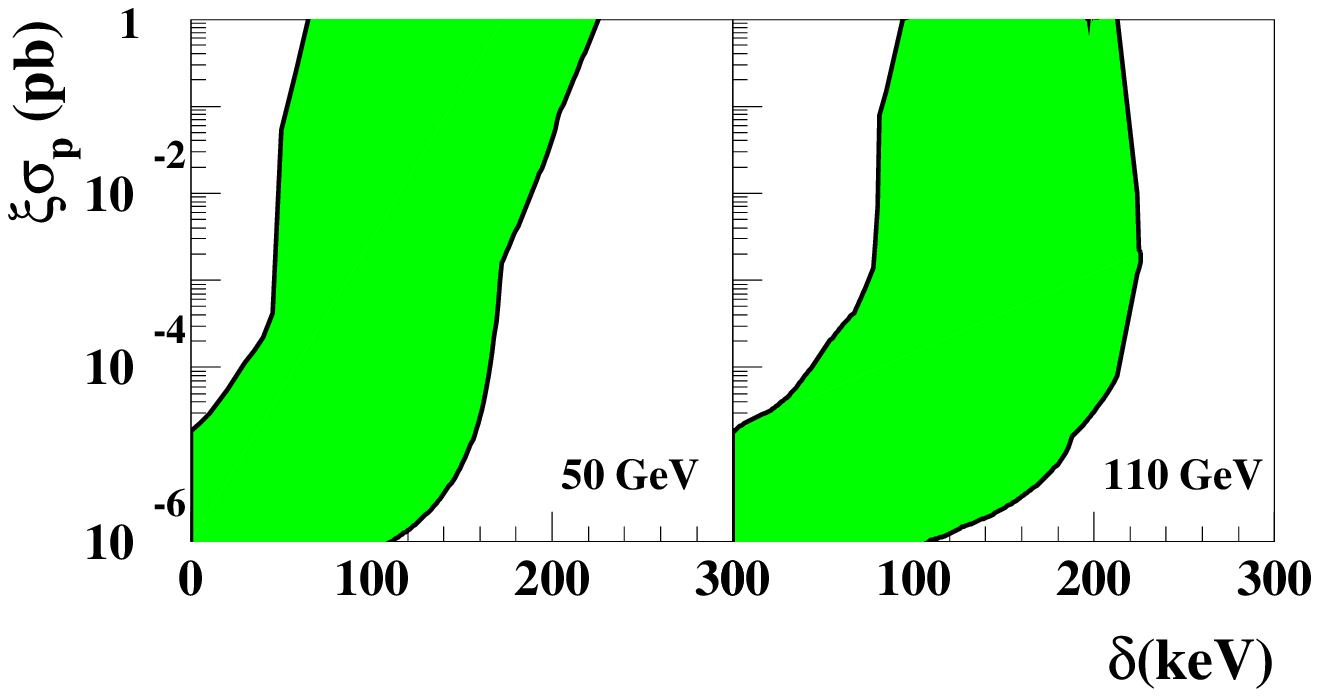}
\end{center}  
\caption{
{\it Case of a WIMP with {\em preferred inelastic} interaction
in the given model frameworks.} Examples of
slices (coloured areas) of the allowed volumes
($\xi \sigma_p$, $\delta$, $m_W$) for some $m_W$ values.
Inclusion of other existing uncertainties on parameters and models 
would further extend the regions; for example,
the use of more favourable form factors 
and different escape velocity
would move them towards lower cross sections.
For details see ref. $^2$.}
\label{fig7}
\end{figure}

Fig. \ref{fig3}, \ref{fig4}, \ref{fig6} and \ref{fig7} show some of the obtained allowed 
regions; details and descriptions of the symbols are given in ref. \cite{RNC}.
The theoretical expectations in the purely SI coupling for the particular case of a neutralino candidate 
in the MSSM with gaugino mass unification at GUT scale released \cite{Bo03} are  
also shown in Fig. \ref{fig5}.

We remind that the allowed regions, we report herein and previously elsewhere,
take into
account the time and energy behaviours of the single-hit experimental data
and have been obtained 
by a maximum likelihood procedure (for a formal description see e.g. refs. \cite{Mod1,Mod2,Mod3}) 
which also determines on this basis 
for each model the constant part of the signal and the background
\footnote{We point out to the attention of the reader that the annual modulation approach 
gives as experimental results the modulation amplitudes, while 
the constant part of the signal, for each considered energy interval,
has to be extracted from the measured counting rate of the single hit events,
which a priori represents the sum of the signal and of possible residual background 
satisfying the trigger condition.
A  natural constraint on the constant part of the signal (and thus on the background) arises 
from the measured upper limits on recoils and has to be accounted as well, as  
done since ref. \cite{Mod3}.}.
In particular, the likelihood function 
itself requires the agreement: i) of the expectations for the modulated part of the signal 
with the measured modulated behaviour for each detector and for each energy bin; ii) 
of the expectations for the unmodulated component of the signal with the respect 
to the measured differential energy distribution and 
- since ref. \cite{Mod3} - also with the 
bound on recoils obtained by pulse shape discrimination 
from the devoted DAMA/NaI-0 data\cite{Psd96}.
The latter one acts in the likelihood procedure as an experimental upper bound on the unmodulated component of the 
signal and -- as a matter of fact -- as an experimental lower bound on the estimate of the background levels. 
Thus, the C.L.'s, we quote for the 
allowed regions, already account for compatibility with the measured differential
energy spectrum and with the measured upper bound on recoils.

Specific arguments on some claimed model dependent comparisons can be found 
in ref.\cite{RNC}. They already account, as a matter of fact, also for the more recent
model dependent CDMS(-II) claim \cite{cdms2} (based on a statistics of 19.4 kg $\cdot$ day 
and on a discrimination technique), where 
DAMA/NaI is not correctly quoted and the more recent 
result of the 7 annual cycles \cite{RNC} is quoted but not accounted for.
In addition, in the particular scenario of ref.\cite{cdms2}, uncertainties from the model (from astrophysics,
 nuclear and particle physics)  as well as some experimental ones,
are not accounted at all and the existing 
interactions and scenarios to which CDMS is largely insensitive -- on the contrary of DAMA/NaI -- are ignored.

\section{Conclusions and perspectives}

DAMA/NaI has been a pioneer experiment running at the 
Gran Sasso National Laboratory of I.N.F.N.
for several years and investigating as first the WIMP annual modulation signature
with suitable sensitivity and control of the running parameters. During seven independent
experiments of one year each one, it has pointed out the presence of a modulation
satisfying the many peculiarities of an effect induced by Dark Matter particles, 
reaching a significant evidence.
As a corollary result, it has also pointed out the complexity of the quest for a
candidate particle mainly because of the present poor knowledge 
on the many astrophysical,
nuclear and particle physics aspects. At present after a devoted R\&D effort, the 
second generation DAMA/LIBRA (a $\simeq$250 kg more radiopure NaI(Tl) set-up)
has been realised and put in operation since March 2003.
Moreover, a third generation R\&D toward a possible ton 
NaI(Tl) set-up, we proposed in 1996 \cite{IDM96}, is in progress.

\end{document}